\documentstyle[sprocl]{article}
\def\be{\begin{equation}}
\def\ee{\end{equation}}
\def\bea{\begin{eqnarray}}
\def\eea{\end{eqnarray}}

\begin{document}
\title{SYMMETRIES OF THE STANDARD MODEL\\ 
WITHOUT AND WITH A RIGHT-HANDED NEUTRINO}
\author{ERNEST MA}
\address{Department of Physics, University of California\\
Riverside, CA 92521, USA}
\maketitle\abstracts{Given the particle content of the standard model 
without and with a right-handed neutrino, the requirement that all anomalies 
cancel singles out a set of possible global symmetries which can be gauged. 
I review this topic and propose a new gauge symmetry $B - 3 L_\tau$ in the 
context of the minimal standard model consisting of the usual three families 
of quarks and leptons plus just one $\nu_R$.  The many interesting 
phenomenological consequences of this hypothesis are briefly discussed.}

\section{Cancellation of Anomalies}

In the minimal standard model, under the gauge group $SU(3)_C \times SU(2)_L 
\times U(1)_Y$, the quarks and leptons transform as:
\begin{equation}
\left[ \begin{array} {c} u \\ d \end{array} \right]_L \sim (3, 2, 1/6), 
~~~ u_R \sim (3, 1, 2/3), ~~~ d_R \sim (3, 1, -1/3);
\end{equation}
\begin{equation}
\left[ \begin{array} {c} \nu \\ l \end{array} \right]_L  \sim (1, 2, -1/2), 
~~~ l_R \sim (1, 1, -1).
\end{equation}
The sum of the U(1) axial charges is
\begin{equation}
6 (1/6) + 2 (-1/2) - 3 (2/3) - 3 (-1/3) - (-1) = 0.
\end{equation}
This means that the mixed gauge-gravitational anomaly~\cite{1} is absent. 
The number of SU(2) fermion doublets is even.  Hence the SU(2) global 
anomaly~\cite{2} is also absent.

Because left-handed and right-handed fermions transform differently, there 
are potentially several axial-vector triangle anomalies~\cite{3} in the 
standard model.  However, they are all canceled as follows:
\begin{equation}
[SU(3)]^2 U(1): ~~~ 2(1/6) - (2/3) - (-1/3) = 0,
\end{equation}
\begin{equation}
[SU(2)]^2 U(1): ~~~ 3(1/6) + (-1/2) = 0,
\end{equation}
\begin{equation}
[U(1)]^3: ~~~ 6(1/6)^3 + 2(-1/2)^3 - 3(2/3)^3 - 3(-1/3)^3 - (-1)^3 = 0.
\end{equation}

\section{Global and Gaugeable Symmetries}

There are four global symmetries in the minimal standard model, corresponding 
to baryon number $B$ and the three lepton numbers $L_e$, $L_\mu$, and 
$L_\tau$.  Each current [{\it i.e.} $J_B^\alpha = \sum_i (\bar q_{iL} 
\gamma^\alpha q_{iL} + \bar q_{iR} \gamma^\alpha q_{iR})$, etc.] is 
conserved at the classical (tree) level.  However, at the quantum level 
[{\it i.e.} when loops are included], each current by itself is not 
conserved.  In fact, it was shown long ago~\cite{4} that
\begin{equation}
\partial_\alpha J^\alpha_{B+L} = {g^2 \over {16 \pi^2}} N_f W_i^{\mu \nu} 
\tilde W_{i \mu \nu},
\end{equation}
where $W_i^{\mu \nu}$ is the $SU(2)_L$ field tensor and $\tilde W$ its dual, 
$g$ is the corresponding gauge coupling, and $N_f$ is the number of families 
of quarks and leptons.  Nonperturbative pseudo-particle contributions to the 
above result in the selection rule
\begin{equation}
\Delta (B + L) = 2 N_f \nu,
\end{equation}
where
\begin{equation}
\nu = {g^2 \over {32 \pi^2}} \int d^4 x W_i^{\mu \nu} \tilde W_{i \mu \nu} = 
\pm 1, \pm 2, ...
\end{equation}
Note that
\begin{equation}
\partial_\alpha J^\alpha_{B-L} \propto 3(3)(1/3) - 1 - 1 - 1 = 0;
\end{equation}
hence the electroweak phase transition conserves $B-L$.  However, the chiral 
$B-L$ anomaly is nonzero:
\begin{equation}
[U(1)_{B-L}]^3: ~~~ 6(1/3)^3 + 2(-1)^3 - 6(1/3)^3 - (-1)^3 \neq 0;
\end{equation}
Hence $B-L$ itself cannot be gauged without the addition of one $\nu_R$ per 
family.

It is already well-known that if there is one $\nu_R$ per family, then the 
standard model may be extended to become a left-right symmetric model with
\begin{equation}
Q = T_{3L} + Y = T_{3L} + T_{3R} + {1 \over 2} (B - L).
\end{equation}
In other words, $SU(2)_L \times U(1)_Y$ naturally becomes $SU(2)_L \times 
SU(2)_R \times U(1)_{B-L}$.

Without $\nu_R$, there are still three possible symmetries which may be 
gauged~\cite{5}: $L_e - L_\mu$, $L_e - L_\tau$, or $L_\mu - L_\tau$. 
The $L_i - L_j$ anomalies cancel because
\begin{equation}
[SU(2)]^2 (L_i - L_j): ~~~ 1 - 1 = 0,
\end{equation}
\begin{equation}
[U(1)_Y]^2 (L_i - L_j): ~~~ [2(-1/2)^2 - (-1)^2](1-1) = 0,
\end{equation}
\begin{equation}
U(1)_Y (L_i - L_j)^2: ~~~ 2(-1/2)(1)^2 - (-1)(1)^2 + 2(-1/2)(-1)^2 - 
(-1)(-1)^2 = 0,
\end{equation}
\begin{equation}
(L_i - L_j)^3: ~~~ 2(1)^3 - (1)^3 + 2(-1)^3 - (-1)^3 = 0.
\end{equation}
The $Z'$ boson associated with this U(1) may show up in 
$e^+ e^-$ or $\mu^+ \mu^-$ collisions as a resonance decaying into 
charged-lepton pairs.

Without $\nu_R$, all neutrinos are massless, protected respectively by the 
additive lepton numbers $L_e$, $L_\mu$, $L_\tau$, unless a Higgs triplet is 
added~\cite{6}.  In general, there may be a certain number of $\nu_R$'s, 
not necessarily equal to the number of $\nu_L$'s, as discussed for example in 
Ref.~7.

\section{Standard Model + one $\nu_R$}

If there is only one $\nu_R \sim (1, 1, 0)$, then it can couple only to one 
linear combination of $\nu_{iL}$.  Assuming that its Majorana mass is $M_R$, 
the $4 \times 4$ mass matrix spanning $\nu_{iL}$ and $\nu^c_R$ is given by
\begin{equation}
{\cal M} = \left[ \begin{array} {c@{\quad}c@{\quad}c@{\quad}c} 0 & 0 & 0 & m_1 
\\ 0 & 0 & 0 & m_2 \\ 0 & 0 & 0 & m_3 \\ m_1 & m_2 & m_3 & M_R \end{array} 
\right].
\end{equation}
If $m_{1,2,3} << M_R$, one linear combination of $\nu_{iL}$ gets a seesaw 
mass~\cite{8}
\begin{equation}
m_{\nu_0} \sim {{m_1^2 + m_2^2 + m_3^2} \over M_R}.
\end{equation}
At this level, there are two massless neutrinos left, but since no lepton 
number remains conserved, they must pick up finite radiative masses. 
Specifically, they do so in two loops through the exchange of two $W$ 
bosons~\cite{9}.  Their masses are proportional to $m_{\nu_0}$ of Eq.~(18) 
and are functions of the charged-lepton masses with double GIM 
suppression~\cite{10}.  A detailed analytical and numerical study of this 
mechanism has been made~\cite{11}.

With one $\nu_R$, the global symmetries left are additive baryon number $B$ 
and multiplicative lepton number $L$.  This is true also in the canonical 
scenario where three $\nu_R$'s are added to obtain three seesaw neutrino 
masses.

\section{Gauged $B - 3 L_\tau$}

As mentioned already, $B$ cannot be gauged by itself without adding to 
the particle content of the standard model.  Nevertheless, the phenomenology 
of such a possibility has been discussed~\cite{12}.  On the other hand, 
if the one $\nu_R$ added to the minimal standard model is required to be 
$\nu_{\tau R}$, then $B - 3 L_\tau$ can be gauged~\cite{13}.  The various 
axial-vector triangle anomalies are canceled as follows:
\begin{equation}
[SU(2)]^2 U(1)_X: ~~~ 3(3)(1/3) - 3 = 0,
\end{equation}
\begin{equation}
[U(1)_X]^2 U(1)_Y: ~~~ (1/3)^2[2(1/6) - (2/3) - (-1/3)] + (-3)^2[2(-1/2) - 
(-1)] = 0,
\end{equation}
\begin{eqnarray}
[U(1)_Y]^2 U(1)_X: &~& 3(3)[2(1/6)^2 - (2/3)^2 - (-1/3)^2](1/3) \nonumber \\ 
&~& + [2(-1/2)^2 - (-1)^2](-3) = 0.
\end{eqnarray}

Where does $B - 3 L_\tau$ come from?  First consider $B - L$ for each family. 
In that case, the standard model can be extended to $SU(4) \times SU(2)_L 
\times U(1)'$, under which the quarks and leptons transform as:
\begin{equation}
\left[ \begin{array} {c@{\quad}c} u & \nu \\ d & l \end{array} \right]_L 
\sim (4,2,0), ~~ [u~\nu]_R \sim (4,1,1/2), ~~ [d~l]_R \sim (4,1,-1/2),
\end{equation}
with the electric charge given by
\begin{equation}
Q = {1 \over 2} (B-L) + T_{3L} + Y'.
\end{equation}
Since $Y'$ takes on the values $\pm 1/2$, it is natural to identify $U(1)'$ 
as $U(1)_R$ and extend~\cite{14} the gauge group to $SU(4) \times SU(2)_L 
\times SU(2)_R$.  In the case of $B - 3 L_\tau$, consider instead 
$SU(10) \times SU(2)_L \times U(1)'$, with
\begin{equation}
Q = {1 \over 5} (B - 3 L_\tau) + T_{3L} + Y'.
\end{equation}
The quarks and leptons are now organized as follows.
\begin{equation}
\left[ \begin{array} {c@{\quad}c@{\quad}c@{\quad}c} u & c & t & \nu_\tau \\ 
d & s & b & \tau \end{array} \right]_L \sim (10,2,1/10), ~~ \left[ 
\begin{array} {c} \nu_e \\ e \end{array} \right]_L, \left[ \begin{array} 
{c} \nu_\mu \\ \mu \end{array} \right]_L \sim (1,2,-1/2),
\end{equation}
\begin{equation}
[u~c~t~\nu_\tau]_R \sim (10,1,3/5), ~~ [d~s~b~\tau]_R \sim (10,1,-2/5), ~~ 
e_R,~\mu_R \sim (1,1,-1).
\end{equation}
It is clear that $SU(10)$ breaks down to $SU(9) \times U(1)_{B - 3 L_\tau}$, 
and $SU(9)$ to $[SU(3)]^3 \times U(1) \times U(1)$.  The possibility of 
$[SU(3)]^3$ has been discussed~\cite{15}, but the two extra U(1) symmetries 
remain unexplored.  They may be relevant in understanding whether and how 
families of quarks could be different.  Note also that $L_e - L_\mu$ is 
still gaugeable together with $B - 3 L_\tau$.

The essential prediction of gauged $B - 3 L_\tau$ is of course a new gauge 
boson.  Let me call it $X$.  Since it does not couple to $e$ or $\mu$ or their 
corresponding neutrinos, there is no direct phenomenological constraint from 
the best known high-energy physics experiments, such as $e^+ e^-$ 
annihilation, deep-inelastic scattering of $e$ or $\mu$ or $\nu_\mu$ on 
nuclei, or the observation of $e^+ e^-$ or $\mu^+ \mu^-$ pairs in hadronic 
collisions.  Although $X$ does contribute to purely hadronic interactions, 
its presence is effectively masked by the enormous background due to quantum 
chromodynamics (QCD).  However, unlike the case of a gauge boson coupled only 
to baryon number~\cite{12}, $X$ also couples to $L_\tau$.  Assuming that 
$\nu_{\tau R}$ and the $t$ quark are too heavy to be decay products of $X$, the 
branching fractions of $X$ are 10/91, 54/91, and 27/91 in the parton 
approximation for the modes $\bar q q$, $\tau^+ \tau^-$, and $\bar \nu_\tau 
\nu_\tau$ respectively.  There are two possible production mechanisms for 
$X$.  It can be directly produced in hadronic collisions: $q \bar q 
\rightarrow X$, or indirectly from $Z$ decay if $M_X < M_Z$.

Consider the sequential decays $Z \rightarrow \bar q q X$, then $X \rightarrow 
\bar \nu_\tau \nu_\tau$, and $Z \rightarrow \bar \nu_\tau \nu_\tau X$, then 
$X \rightarrow \bar q q$.  Since the neutrinos are undetected, these 
processes have the same effective final states as in the search for an 
invisible Higgs boson in $Z$ decay~\cite{16}.  Another decay mode to 
search for is $Z \rightarrow \tau^+ \tau^- X$, then $X \rightarrow \tau^+ 
\tau^-$.  From the present experimental nonobservation of the above, 
the coupling $g_X$ has to be smaller than about 0.2 for $M_X < 50$ GeV. 
From the observed $e - \mu - \tau$ universality in $Z$ decay, a limit of 
$g_X$ less than 0.2 to 0.4 is also obtained for $M_X$ between 50 and 150 
GeV.  Details will be presented elsewhere~\cite{17}.

The global symmetries left up to now are additive $B$, multiplicative $L_\tau$ 
[from the breaking of gauged $B - 3 L_\tau$ with $\chi^0 \sim (1,1,0;6)$], 
and additive $L_e$, $L_\mu$.  Whereas $\nu_\tau$ gets a seesaw mass, 
$\nu_e$ and $\nu_\mu$ remain massless.  To allow them nonzero masses, 
a second Higgs doublet is added:
\begin{equation}
\left[ \begin{array} {c} \eta^+ \\ \eta^0 \end{array} \right] \sim (1, 2, 
1/2; -3),
\end{equation}
which induces mixing between $\tau$ and the other two leptons.  The 
$3 \times 3$ charged-lepton mass matrix is now given by
\begin{equation}
{\cal M}_l = \left[ \begin{array} {c@{\quad}c@{\quad}c} m_e & 0 & 0 \\ 
0 & m_\mu & 0 \\ a_e & a_\mu & m_\tau \end{array} \right].
\end{equation}
There are some additional consequences.  (1) The $X$ boson mixes with $Z$, 
requiring the vacuum expectation value of $\eta^0$ to be much smaller than 
those of $\phi^0$ and $\chi^0$ for it to be consistent with data.  (2) Only 
one multiplicative lepton number is left.  (3) Two neutrinos acquire radiative 
masses through one-loop $\eta^0$ and two-loop $W$ exchange, but the values 
so obtained are too small to be of phenomenological interest.  The addition 
of a charged scalar singlet $\chi^- \sim (1, 1, -1; -3)$ allows it to couple 
to $\nu_l \tau_L - l_L \nu_\tau$, hence the exchange and mixing of $\chi^-$ 
with the physical linear combination formed by $\phi^-$ and $\eta^-$ will be 
able to generate~\cite{18} realistic one-loop neutrino masses.  (4) Lepton 
family number is violated by the parameters $a_e$ and $a_\mu$.  The best 
present limits are from $\mu - e$ conversion in nuclei:
\begin{equation}
{g_X^2 \over M_X^2} a_\mu a_e < 1.2 \times 10^{-12};
\end{equation}
from the decay of $\tau$ to $\mu \pi^+ \pi^-$ over $\nu \pi^- \pi^0$:
\begin{equation}
{g_X^2 \over M_X^2} a_\mu m_\tau < 3.8 \times 10^{-7};
\end{equation}
and from the decay of $\tau$ to $e \pi^+ \pi^-$ over $\nu \pi^- \pi^0$:
\begin{equation}
{g_X^2 \over M_X^2} a_e m_\tau < 2.9 \times 10^{-7}.
\end{equation}
The rare decay $K^+ \rightarrow \pi^+ \nu_\tau \bar \nu_\tau$ gets a 
contribution from $X$, but since the $X \bar q q$ interaction is vectorial, 
it is like that of a heavy photon or gluon, hence it is very much suppressed 
relative to the $Z$ contribution which has axial-vector couplings to quarks.

\section{Conclusion and Outlook}

The number of $\nu_R$'s (0, 1, 2, 3, ...) which may accompany the minimal 
standard model is crucial to understanding the symmetries of the (extended) 
model and its leptonic properties.  Both neutrino physics and high-energy 
accelerator physics are on the verge of possible major discoveries.  The next 
several years will be decisive in leading us forward in their theoretical 
understanding, and may even discover radically new physics beyond the 
standard model.

\section*{Acknowledgments} I thank Rajiv Gavai and all the other organizers 
for their great hospitality and a stimulating workshop.  This work 
was supported in part by the U.~S.~Department of Energy under Grant 
No.~DE-FG03-94ER40837.

\end{document}